\documentclass[sigconf]{acmart}
\usepackage{array}
\usepackage{graphicx}
\usepackage{clrscode}
\usepackage{balance}
\usepackage{subfigure}
\usepackage{multirow}
\usepackage{float}
\usepackage{color}
\usepackage{pmat}
\usepackage{amsopn}
\usepackage{mathrsfs}
\usepackage{mathtools}
\usepackage{amsmath}
\usepackage{arydshln}
\usepackage{hyperref}
\usepackage{multicol}
\usepackage{blkarray}
\usepackage{enumerate}
\usepackage{courier}
\usepackage{rotating}
\usepackage{booktabs}
\usepackage{algorithm}
\usepackage{algorithmic}

\copyrightyear{2018}
\acmYear{2018} 
\setcopyright{acmcopyright}

\acmConference{SIGIR'18}{}{July 8 -- 12, 2018, Ann Arbor, Michigan, U.S.A}
\acmPrice{15.00}

\begin{document}

\title{\mbox{Learning over Knowledge-Base Embeddings for Recommendation}}

%

\author{Yongfeng Zhang$^{1}$, Qingyao Ai$^{2}$, Xu Chen$^3$, Pengfei Wang$^4$}
\affiliation{
 \institution{$^1$Rutgers University $^2$ University of Massachusetts Amherst}
  \institution{$^3$Tsinghua University $^4$Beijing University of Posts and Telecommunications}
}
\email{yongfeng.zhang@rutgers.edu, aiqy@cs.umass.edu, xu-ch14@mails.tsinghua.edu.cn, wangpengfei@bupt.edu.cn}

\begin{abstract}
State-of-the-art recommendation algorithms -- especially the collaborative filtering (CF) based approaches with shallow or deep models -- usually work with various unstructured information sources for recommendation, such as textual reviews, visual images, and various implicit or explicit feedbacks. Though structured knowledge bases were considered in content-based approaches, they have been largely neglected recently due to the availability of vast amount of data, and the learning power of many complex models.

However, structured knowledge bases exhibit unique advantages in personalized recommendation systems. When the explicit knowledge about users and items is considered for recommendation, the system could provide highly customized recommendations based on users' historical behaviors.
A great challenge for using knowledge bases for recommendation is how to integrated large-scale structured and unstructured data, while taking advantage of collaborative filtering for highly accurate performance. Recent achievements on knowledge base embedding sheds light on this problem, which makes it possible to learn user and item representations while preserving the structure of their relationship with external knowledge. In this work, we propose to reason over knowledge base embeddings for personalized recommendation. Specifically, we propose a knowledge base representation learning approach to embed heterogeneous entities for recommendation. Experimental results on real-world dataset verified the superior performance of our approach compared with state-of-the-art baselines.
\end{abstract}


\maketitle

\section{Introduction}
Most of the existing Collaborative filtering (CF)-based recommendation systems work with various unstructured data such as ratings, reviews, or images to profile the users for personalized recommendation. Though effective, it is difficult for existing approaches to model the explicit relationship between different information that we know about users and items. 
In this paper, we would like to ask a key question, i.e., \emph{``can we extend the power of collaborative filtering upon large-scale structured user behavior data?''}. 
The main challenge to answer this question is how to effectively integrate different types of user behaviors and item properties, while preserving the internal relationship betwen them to enhance the final performance of personalized recommendation.

Fortunately, the emerging success on knowledge base embeddings may shed some light on this problem, where heterogeneous knowledge entities and relations can be projected into a unified embedding space. By encoding the rich information from multi-type user behaviors and item properties into the final user/item embeddings, we can enhance the recommendation performance while preserving the internal structure of the knowledge.

Inspired by the above motivation, in this paper, we design a novel collaborative filtering framework over knowledge graph. The main building block is an integration of traditional CF and knowledge-base embedding technology. More specifically, we first define the concept of user-item knowledge graph, which encodes our knowledge about the users and items as a relational graph structure. The user-item knowledge graph focuses on how to depict different types of user behaviors and item properties over heterogenous entities and relations in a unified framework. Then, we extend the design philosophy of collaborative filtering (CF) to learn over the knowledge graph for personalized recommendation.

\begin{figure}[t!]
\centering
\setlength{\fboxrule}{0.pt}
\setlength{\fboxsep}{0.pt}
\fbox{
\includegraphics[width=0.95\linewidth, ]{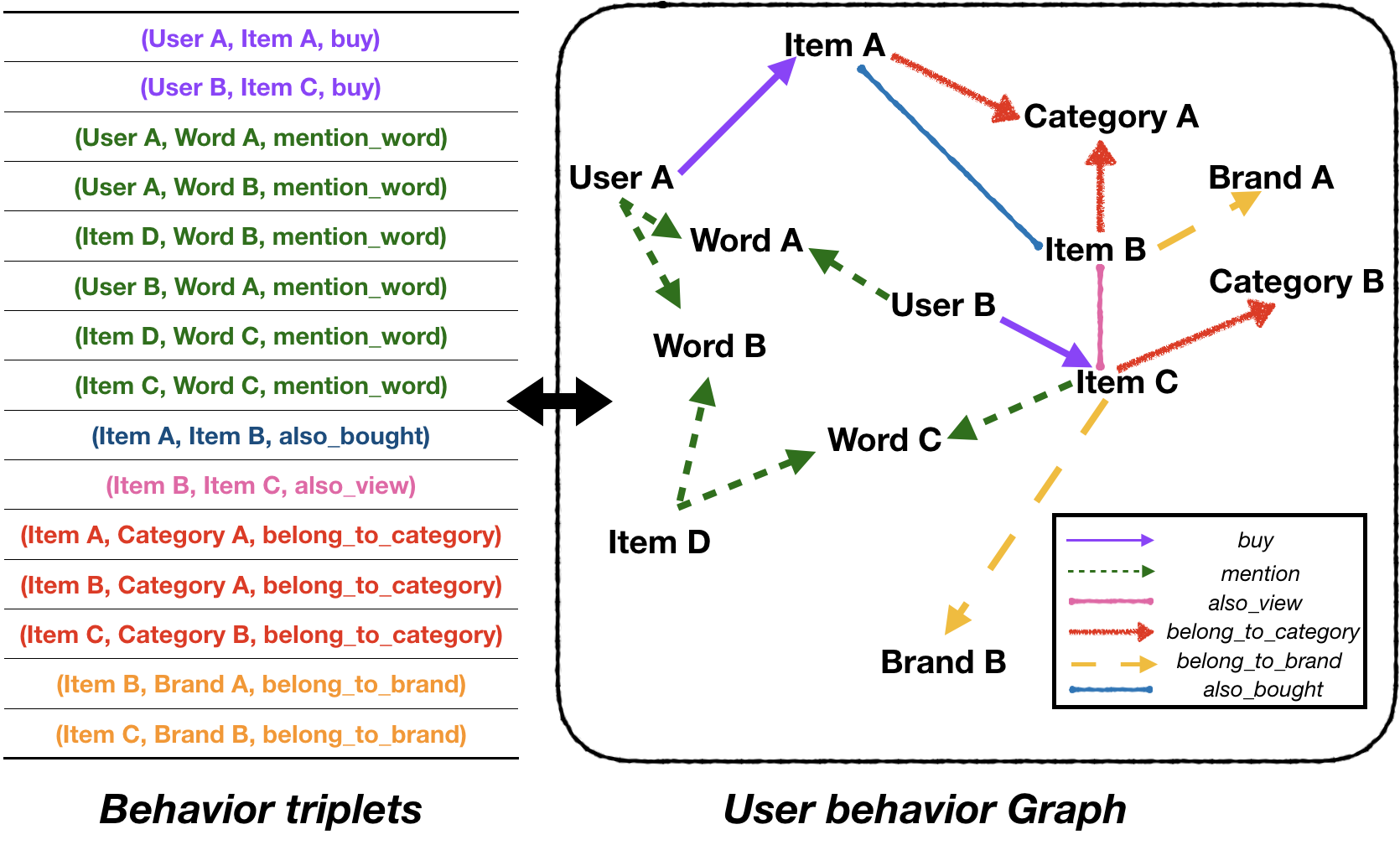}
}
\vspace{-10pt}
\caption{A toy example of user-item knowledge graph. In the left is a set of triplets of user behaviors and item properties, and in the right is the corresponding  graph structure.}
\vspace{-15pt}
\label{example}
\end{figure}

\noindent
\textbf{Contributions.} The main contributions of this paper can be summarized as follows:

$\bullet$ We propose to directly reason over a structured knowledge-base with embeddings for recommendation.

$\bullet$ We extend traditional collaborative filtering to learn over the heterogenous knowledge-base embeddings, which makes it possible to capture the user preferences more comprehensively.

$\bullet$ Extensive experiments verify that our model can consistently outperform many state-of-the-art baselines on real-world e-commerce datasets.

In the following part of the paper, we first illustrate our models in section 2, and verify its effectiveness with experimental results in section 3. At last, the related work and conclusions are presented in section 4 and 5, respectively.

\section{Collaborative Filtering with Knowledge-Graph}
In this section, we illustrate our model to make recommendations by learning knowledge-graph embeddings with collaborative filtering. 

\subsection{Model structure}
Incorporating additional information has been a widely used method to enhance recommendation performance, however, the objects in a practical recommender system are usually very heterogeneous (\emph{e.g.} product brands, categories, and user reviews, etc), and the relations between them can also be quite different (\emph{e.g.} belong to, buy, view, etc).
To tackle with such heterogenous data in a unified framework, we first define a user-item knowledge-graph structure specialized for recommender systems, then we conduct collaborative filtering on this graph to provide personalized recommendations.  

\subsubsection{User-Item Knowledge Graph in Recommender system}
In the context of recommender system, the user-item knowledge graph is built from a set of triplets. Each triplet $(i,j,r)$ is composed of a head entity $i$, a tail entity $j$, and the relation $r$ from $i$ to $j$. The semantic for a triplet $(i,j,r)$ is that \emph{$i$ has a (directed) relation $r$ with $j$}. For example, $(user, item, buy)$ means that the \emph{$user$} has \emph{bought} the \emph{$item$} before, and $(item, brand, belong\_to)$ means that the $item$ \textit{belongs to} a particular $brand$.

Specifically, we define 5 types of entities and 6 types of relations in our system, where the entities include \textit{user, item, word, brand} and \textit{category}, 
while the relations include:\\
$\bullet$ $buy$: the relation from a user to an item, meaning that the user has bought the item.\\
$\bullet$ $belong\_to\_category$: the relation from an item to a category, meaning that the item belongs to the category.\\
$\bullet$ $belong\_to\_brand$: the relation from an item to a brand, meaning that the item belongs to the brand.\\
$\bullet$ $mention\_word$: relation from a user or an  item to a word, meaning that the word is mentioned in the reviews of the user or item.\\
$\bullet$ $also\_bought$: the relation from an item to another item, meaning that users who bought the first item also bought the second item.\\
$\bullet$ $also\_view$: the relation from an item to another item, meaning that users who bought the first item also viewed the second item.\\

An example of the constructed user-item knowledge graph can be seen in Figure~\ref{example}.

\begin{algorithm}[t]
\caption{Collaborative Filtering based on Knowledge Graph}
\label{alg:A}
\begin{algorithmic}
\REQUIRE Entity set $E$, relation set $R$, triplet set $S$, dimension $D$, number of negative samples $k$;\\
\STATE Randomly initialize the embeddings for $r\in R$ and $e\in E$
\FOR {$(e_i,e_j,r)$ in $S$}
\STATE $l \leftarrow 0, S^t \leftarrow \varnothing, S^h \leftarrow \varnothing$
\REPEAT
\STATE $e_j^\prime \gets sample(E)$, $e_i^\prime \gets sample(E)$  
\STATE $S^t \gets S^t \cup (e_i,e_j^\prime,r), S^h \gets S^h \cup (e_i^\prime,e_j,r)$ 
\STATE $l \gets l + 1$
\UNTIL{$l=k$} 
\STATE Update embeddings according to $\nabla L$
\ENDFOR
\ENSURE Embeddings of the subjects and relations
\end{algorithmic}
\end{algorithm}


\begin{table*}[t]
\small
\caption{Performance on top-10 recommendation between the baselines and our model (all the values in the table are percentage numbers with `\%' omitted), where the bolded numbers indicate the best performance of each column.
The first block shows the results of the baselines, where the stared numbers indicate the best baseline performances; 
the the second line from the bottom presents the results of our model. 
The last line shows the percentage increment of our results against the best baseline (i.e., JRL), which are significant at \textit{\textbf{p}}=0.001.}
\centering
\vspace{-5pt}
\setlength{\tabcolsep}{2pt}
\begin{tabular}
	{l|rrrr|rrrr|rrrr|rrrr} \hline\hline
	Dataset &  \multicolumn{4}{c|}{CDs} & \multicolumn{4}{c|}{Clothing} & \multicolumn{4}{c|}{Cell Phones} & \multicolumn{4}{c}{Beauty}\\\hline
	Measures(\%) & NDCG & Recall & HT & Prec & NDCG & Recall & HT & Prec & NDCG & Recall & HT & Prec & NDCG & Recall & HT & Prec \\\hline
BPR  & 2.009 & 2.679 & 8.554 & 1.085 & 0.601 & 1.046 & 1.767 & 0.185 & 1.998 & 3.258 & 5.273 & 0.595 & 2.753 & 4.241 & 8.241 & 1.143\\
BPR\_HFT & 2.661 & 3.570 & 9.926 & 1.268 & 1.067 & 1.819 & 2.872 & 0.297 & 3.151 & 5.307 & 8.125 & 0.860 & 2.934  & 4.459 & 8.268 & 1.132\\
VBPR & 0.631 & 0.845 & 2.930 & 0.328 & 0.560 & 0.968 & 1.557 & 0.166 & 1.797 & 3.489 & 5.002 & 0.507 & 1.901 & 2.786 & 5.961 & 0.902\\
DeepCoNN  & 4.218 & 6.001 & 13.857 & 1.681 & 1.310 & 2.332 & 3.286 & 0.229 & 3.636 & 6.353 & 9.913 & 0.999 & 3.359 & 5.429 & 9.807 & 1.200\\
CKE  & 4.620 & 6.483 & 14.541 & 1.779 & 1.502 & 2.509 & 4.275 & 0.388 & 3.995 & 7.005 & 10.809 & 1.070 & 3.717 & 5.938 & 11.043 & 1.371\\
JRL & {5.378}$^*$ & {7.545}$^*$ & {16.774}$^*$ & {2.085}$^*$ & {1.735}$^*$ & {2.989}$^*$ & {4.634}$^*$ & {0.442}$^*$ & {4.364}$^*$ & {7.510}$^*$ & {10.940}$^*$ & {1.096}$^*$ & {4.396}$^*$ & {6.949}$^*$ & {12.776}$^*$ & {1.546}$^*$\\\hline
CFKG & \textbf{5.563} & \textbf{7.949} & \textbf{17.556} & \textbf{2.192} & \textbf{3.091} & \textbf{5.466} & \textbf{7.972} & \textbf{0.763} & \textbf{5.370} & \textbf{9.498} & \textbf{13.455} & \textbf{1.325} & \textbf{6.370} & \textbf{10.341} & \textbf{17.131} & \textbf{1.959}\\\hline
Improvement & 3.44 & 5.35  & 4.66 & 5.13 & 78.16 & 82.87 & 72.03 & 72.62 & 23.05 & 26.47 & 22.99 & 20.89 & 44.90 & 48.81 & 34.09 & 26.71\\\hline
\end{tabular}\label{tab:result}
\vspace{-5pt}
\end{table*}

\subsubsection{Collaborative Filtering based on User-Item Knowledge Graph}
The user-item knowledge graph provides us with the ability to access different information sources and multi-type behaviors in a unified manner. In this section, we conduct collaborative filtering on this graph for accurate user profiling and personalized recommendation. Inspired by ~\cite{bordes2013translating}, we project each entity and relation into a unified low-dimensional embedding space. Intuitively, the embedding of a tail entity should be close to its translated head entity embedding. Formally, for a triplet $(i,j,r)$, suppose the embeddings of $i,j,r$ are $e_i,e_j,e_r$, respectively, then we want that $trans_{e_r}(e_i) \approx e_j$. 
Considering all the observed triplets $S$, we minimize a margin-based loss to learn the embeddings as follows:
\begin{equation}
\begin{aligned}
L &= \sum_{(i,j,r)\in S} \Big\{\sum_{(i,j^\prime,r)\in S^t} \big[\gamma+d\big(trans_{e_r}(e_i), e_j\big)-d\big(trans_{e_r}(e_{i}), e_{j^\prime}\big)\big]_+ \\
&+ \sum_{(i^\prime,j,r)\in S^h} \big[\gamma+d\big(trans_{e_r}(e_i), e_j\big)-d\big(trans_{e_r}(e_{i^\prime}), e_{j}\big)\big]_+ \Big\}
\end{aligned}
\end{equation}
where, $S^t$ is the set of negative triplets that replace the tail by a random entity, and $T^h$ is another set of negative triplets that replace the head by a random entity. $d(\cdot)$ is a metric function to measure the distance between two embeddings, where we select $\ell_2$-norm as its specific implementation. 
$trans_{e_r}(e_{i})$ is an arbitrary translation function, or even a neural network, for here, we adopt the addition function $trans_{e_r}(e_{i}) = e_r + e_i$ as in the transE model \cite{bordes2013translating}, because it gives us better efficiency and effectiveness on our dataset. However, it is not necessarily restricted to this function and many other choices can be used in practice.

In the loss function $L$, we essentially try to discriminate the observed triplets from the corrupted ones by a hinge loss function, and the embeddings will be forced to recover the ground truth.
Our model can be learned by stochastic gradient descent (SGD), and the model learning algorithm is shown in Algorithm~\ref{alg:A}.

\subsubsection{Personalized Recommendation}
We will obtain the embeddings for all entities and relations in the graph once our model is optimized. To generate personalized recommendations for a particular user, we take advantage of the relation type \textit{buy}. Specifically, suppose the embedding of the relation $buy$ is $e_{buy}$, and the embedding of a target user is $e_u$, then we can generate recommendations for the user by ranking the candidate items $e_j$ in ascending order of the distance $d(trans_{e_{buy}}(e_i), e_j)$.

\section{Experiments}
In this section, we evaluate our proposed models by comparing with many state-of-the-art methods. We begin by introducing the experimental setup, and then analyze the experimental results.

\subsection{Experimental Setup}\label{set}

\noindent
\textbf{Datasets}. Experiments are conducted on the Amazon e-commerce dataset~\cite{he2016ups}. We adopt four representative sub-datasets in terms of size and sparsity, which are CD, Clothing, Cell Phone, and Beauty. Statistics of the four datasets are summarized in Table~\ref{tb-dataset}.

\begin{table}[!b]
\vspace{-10pt}
\small
\centering
\vspace{-10pt}
\caption{Statistics of the datasets.}
\vspace{-10pt}
\begin{tabular}{p{2cm}<{\centering}|p{0.8cm}<{\centering}|p{0.8cm}<{\centering}|p{1.6cm}<{\centering}|p{1cm}<{\centering}}
\hline\hline
       Datasets               &\#Users   &\#Items&\#Interactions &Density \\ \hline
{CDs}&75258&64421&1097592&0.0226\%\\\hline
{Clothing}&39387&23033&278677&0.0307\% \\\hline
{Cell Phones}&27879&10429&194493&0.0669\% \\\hline
{Beauty}&22363&12101&198502&0.0734\% \\\hline
\end{tabular}
\label{tb-dataset}
\end{table}

\noindent
\textbf{Evaluation methods}. In our experiments, we leverage the widely used Top-N recommendation measurements including \textbf{Precision}, \textbf{Recall}, \textbf{Hit-Ratio} and \textbf{NDCG} to evaluate our model as well as the baselines. The former three measures aim to evaluate the recommendation quality without considering the ranking positions, while the last one evaluates not only the accuracy but also the ranking positions of the correct items in the final list.

\noindent
\textbf{Baselines}. We adopt the following representative and state-of-the-art methods as baselines for performance comparison:

$-$ \textbf{BPR: } The bayesian personalized ranking~\cite{bpr} model is a popular method for top-N recommendation. We adopt matrix factorization as the prediction component for BPR.

$-$ \textbf{BPR\_HFT: } The hidden factors and topics model~\cite{mcauley2013hidden} is a recommendation method leveraging textual reviews, however, the original model was designed for rating prediction rather than top-N recommendation. To improve its performance, we learn HFT under the BPR pair-wise ranking framework for fair comparison.

$-$ \textbf{VBPR: } The visual bayesian personalized ranking~\cite{he2016vbpr} model is a state-of-the-art method for recommendation with images.

$-$ \textbf{DeepCoNN: } A review-based deep recommender~\cite{zheng2017joint}, which leverages convolutional neural network (CNN) to jointly model the users and items.

$-$ \textbf{CKE: } This is a state-of-the-art neural recommender~\cite{zhang2016collaborativekdd} that integrates textual, visual information, and knowledge base for modeling, but it used knowledge base as regularizers and did not consider the heterogenous connection across different types of entities. 

$-$ \textbf{JRL: } The joint representation learning model~\cite{zhang2017joint} is a state-of-the-art neural recommender, which can leverage multi-model information for Top-N recommendation.


\noindent
\textbf{Parameter settings}. 
All the embedding parameters are randomly initialized in the range of $(0,1)$, and then we update them by conducting stochastic gradient descent (SGD). 
The learning rate is determined in the range of $\{1,0.1,0.01,0.001\}$, and model dimension is tuned in the range of $\{10, 50, 100, 200, 300, 400, 500\}$. This gives us the final learning rate as 0.01 and dimension as 300.
For the baselines, we also determine the final settings by grid search, and for fair comparison, the models designed for rating prediction (i.e. HFT and DeepCoNN) are learned by optimizing the ranking loss similar to BPR. 
When conducting experiments, 70\% items of each user are leveraged for training, while the remaining are used for testing. We generate Top-10 recommendation list for each user in the test dataset.

\begin{table*}[t]
\small
\caption{Performance on top-10 recommendation when incorporating different types of relation in the structured knowledge graph (all the values in the table are percentage numbers with `\%' omitted). The final result (using all relations) is significantly better than all other models (using part of the relations) at p=0.001 level.}
\centering
\vspace{-5pt}
\setlength{\tabcolsep}{2pt}
\begin{tabular}
	{l|rrrr|rrrr|rrrr|rrrr} \hline\hline
	Relations &  \multicolumn{4}{c|}{CDs} & \multicolumn{4}{c|}{Clothing} & \multicolumn{4}{c|}{Cell Phones} & \multicolumn{4}{c}{Beauty}\\\hline
	Measures(\%) & NDCG & Recall & HT & Prec & NDCG & Recall & HT & Prec & NDCG & Recall & HT & Prec & NDCG & Recall & HT & Prec \\\hline
\textit{buy} & 3.822 & 5.185 & 12.828 & 1.628 & 1.019 & 1.754 & 2.780 & 0.265 & 3.387 & 5.806 & 8.548 & 0.848 & 3.658 & 5.727 & 10.549 & 1.305\\
\textit{buy+category} & 4.287 & 5.990 & 14.388 & 1.790 & 1.705 & 3.021 & 4.639 & 0.442 & 3.372 & 5.918 & 8.842 & 0.869 & 3.933 & 6.253 & 11.515 & 1.370\\
\textit{buy+brand} & 3.541 & 4.821 & 12.239 & 1.563 & 1.101 & 1.906 & 2.981 & 0.284 & 3.679 & 6.211 & 9.118 & 0.898 & 4.832 & 7.695 & 13.406 & 1.621\\
\textit{buy+mention} & 4.265 & 5.858 & 13.874 & 1.731 & 1.347 & 2.305 & 3.585 & 0.344 & 4.065 & 7.065 & 10.316 & 1.026 & 4.364 & 6.942 & 12.476 & 1.492\\
\textit{buy+also\_view} & 3.724 & 5.070 & 12.633 & 1.604 & 2.276 & 3.931 & 5.827 & 0.561 & 3.305 & 5.705 & 8.458 & 0.840 & 5.295 & 8.723 & 14.891 & 1.728 \\
\textit{buy+also\_bought} & 5.055 & 7.094 & 16.216 & 2.032 & 1.799 & 3.078 & 4.634 & 0.446 & 5.018 & 8.707 & 12.375 & 1.220 & 5.058 & 8.118 & 13.907 & 1.643 \\\hline
\textit{all} (CFKG) & \textbf{5.563} & \textbf{7.949} & \textbf{17.556} & \textbf{2.192} & \textbf{3.091} & \textbf{5.466} & \textbf{7.972} & \textbf{0.763} & \textbf{5.370} & \textbf{9.498} & \textbf{13.455} & \textbf{1.325} & \textbf{6.370} & \textbf{10.341} & \textbf{17.131} & \textbf{1.959}\\\hline
\end{tabular}\label{tab:info-result}
\vspace{-5pt}
\end{table*}

\subsection{Performance Comparison}
Performance of our Collaborative Filtering with Knowledge Graph (CFKG) model as well as the baseline methods are shown in Table \ref{tab:result}. Basically, the baseline methods can be classified according to the information source(s) used in the method, which are rating-based (BPR), review-based (HFT and DeepCoNN), image-based (VBPR), and heterogenous information modeling (CKE and JRL). Generally, the information sources used by our model include ratings (through the \textit{buy} relation), reviews (through $mention$ relation), and our knowledge about the items (through the \textit{belong\_to\_category, belong\_to\_brand, also\_view} and \textit{also\_bought} relations).

From the experimental results we can see that, both of the review-based models can enhance the performance of personalized recommendation from rating-based methods, and by considering multiple heterogenous information sources, JRL and CKE outperform the other baselines, with JRL achieving the best performance among the baselines. It is encouraging to see that our collaborative filtering with knowledge graph (CFKG) method outperforms the best baseline (JRL) consistently over the four datasets and on all evaluation measures, which verifies the effectiveness of our approach for personalized recommendation. 

However, the performance improvement of our approach may benefit from two potential reasons -- that we used more information sources, and that we used a better structure (i.e., structured knowledge graph) to model heterogenous information. For better understanding, we analyze the contribution of types of relation to our model in the following subsection.

\subsection{Further Analysis on Different Relations}
We experiment the performance of our model when using different relations. Because we eventually need to provide item recommendations for users, the CFKG approach would at least need the \textit{buy} relation to model the user purchase histories. As a result, we test our model when using only the \textit{buy} relation (which simplifies into the translation-based model \cite{he2017translation}), as well as the \textit{buy} relation plus each of the other relations, as shown in Table \ref{tab:info-result}.

We see that when using only the \textit{buy} relation, our CFKG\_\textit{buy} model significantly outperforms the BPR approach on all measures and datasets. On the NDCG measure, the percentage improvement can be 90\% (CDs), 70\% (Clothing), 69\% (Phone), and 33\% (Beauty). Because both BPR and CFKG\_\textit{buy} only used the user purchase information, this observation verifies the effectiveness of using structured knowledge graph embeddings for recommendation. Similarly, CFKG\_\textit{buy+mention} significantly outperforms BPR\_HFT, and also outperforms DeepCoNN except for the recall on the CD dataset. Considering that CFKG\_\textit{buy+mention}, BPR\_HFT and DeepCoNN all work with user purchase history plus textual reviews, this observation further verifies the advantage of using structured knowledge graph for user modeling and recommendation.

Furthermore, we see that adding any one extra relation to the basic \textit{buy} relation gives us improved performance from CFKG\_\textit{buy}. Finally, by modeling all of the heterogenous relation types, the final CFKG model outperforms all baselines and the simplified versions of our model with one or two types of relation, which implies that our knowledge base embedding approach to recommendation is scalable to new relation types, and it has the ability to leverage very heterogeneous information sources in a unified manner.

\section{Related Work}

Using knowledge base to enhance the performance of recommender system is an intuitive idea, which has attracted research attention since very early stages of the recommendation community \cite{trewin2000knowledge,ghani2002building}. However, the difficulty of reasoning over the paths on heterogenous knowledge graphs prevent current approaches from applying collaborative filtering on very different entities and relation types \cite{zhang2016collaborativekdd,catherine2017explainable}, which further makes it difficult to take advantage of the wisdom of crowd. 

Fortunately, recent years have witnessed the success of heterogenous knowledge base embedding techniques \cite{bordes2013translating,wang2014knowledge,lin2015learning}, which can help to learn the embeddings of very different entities to support various application scenarios such as question answering \cite{bordes2014question} and relation extraction from text \cite{lin2015learning}. We believe that learning knowledge base embeddings while preserving the structure of knowledge for reasoning is vital for knowledge-enhanced AI in recommendation systems.



\section{Conclusions and Future Work}\label{sec:conclusions}

In this paper, we propose to learn over heterogenous knowledge base embeddings for personalized recommendation. To do so, we construct the user-item knowledge graph to incorporate both user behaviors and our knowledge about the items. We further learn the knowledge base embeddings with the heterogenous relations collectively, and leverage the user and item embeddings to generate personalized recommendations. Experimental results on real-world datasets verified the superior performance of our approach, as well as its flexibility to incorporate multiple relation types.




\bibliographystyle{ACM-Reference-Format}
\bibliography{paper}


\begin{thebibliography}{00}


\ifx \showCODEN    \undefined \def \showCODEN     #1{\unskip}     \fi
\ifx \showDOI      \undefined \def \showDOI       #1{{\tt DOI:}\penalty0{#1}\ }
  \fi
\ifx \showISBNx    \undefined \def \showISBNx     #1{\unskip}     \fi
\ifx \showISBNxiii \undefined \def \showISBNxiii  #1{\unskip}     \fi
\ifx \showISSN     \undefined \def \showISSN      #1{\unskip}     \fi
\ifx \showLCCN     \undefined \def \showLCCN      #1{\unskip}     \fi
\ifx \shownote     \undefined \def \shownote      #1{#1}          \fi
\ifx \showarticletitle \undefined \def \showarticletitle #1{#1}   \fi
\ifx \showURL      \undefined \def \showURL       #1{#1}          \fi
\providecommand\bibfield[2]{#2}
\providecommand\bibinfo[2]{#2}
\providecommand\natexlab[1]{#1}
\providecommand\showeprint[2][]{arXiv:#2}

\bibitem[\protect\citeauthoryear{Bordes, Chopra, and Weston}{Bordes
  et~al\mbox{.}}{2014}]%
        {bordes2014question}
\bibfield{author}{\bibinfo{person}{A. Bordes}, \bibinfo{person}{S. Chopra},
  {and} \bibinfo{person}{J. Weston}.} \bibinfo{year}{2014}\natexlab{}.
\newblock \showarticletitle{Question Answering with Subgraph Embeddings}. In
  \bibinfo{booktitle}{{\em ACL}}.
\newblock


\bibitem[\protect\citeauthoryear{Bordes, Usunier, Garcia-Duran, Weston, and
  Yakhnenko}{Bordes et~al\mbox{.}}{2013}]%
        {bordes2013translating}
\bibfield{author}{\bibinfo{person}{A. Bordes}, \bibinfo{person}{N. Usunier},
  \bibinfo{person}{A. Garcia-Duran}, \bibinfo{person}{J. Weston}, {and}
  \bibinfo{person}{O. Yakhnenko}.} \bibinfo{year}{2013}\natexlab{}.
\newblock \showarticletitle{Translating embeddings for modeling
  multi-relational data}. In \bibinfo{booktitle}{{\em NIPS}}.
  \bibinfo{pages}{2787--2795}.
\newblock


\bibitem[\protect\citeauthoryear{Catherine, Mazaitis, Eskenazi, and
  Cohen}{Catherine et~al\mbox{.}}{2017}]%
        {catherine2017explainable}
\bibfield{author}{\bibinfo{person}{R. Catherine}, \bibinfo{person}{K.
  Mazaitis}, \bibinfo{person}{M. Eskenazi}, {and} \bibinfo{person}{W. Cohen}.}
  \bibinfo{year}{2017}\natexlab{}.
\newblock \showarticletitle{Explainable Entity-based Recommendations with
  Knowledge Graphs}.
\newblock \bibinfo{journal}{{\em RecSys\/}} (\bibinfo{year}{2017}).
\newblock


\bibitem[\protect\citeauthoryear{Ghani and Fano}{Ghani and Fano}{2002}]%
        {ghani2002building}
\bibfield{author}{\bibinfo{person}{Rayid Ghani} {and} \bibinfo{person}{Andrew
  Fano}.} \bibinfo{year}{2002}\natexlab{}.
\newblock \showarticletitle{Building recommender systems using a knowledge base
  of product semantics}. In \bibinfo{booktitle}{{\em Workshop on Recommendation
  and Personalization in E-Commerce}}. \bibinfo{pages}{27--29}.
\newblock


\bibitem[\protect\citeauthoryear{He, Kang, and McAuley}{He
  et~al\mbox{.}}{2017}]%
        {he2017translation}
\bibfield{author}{\bibinfo{person}{Ruining He}, \bibinfo{person}{Wang-Cheng
  Kang}, {and} \bibinfo{person}{Julian McAuley}.}
  \bibinfo{year}{2017}\natexlab{}.
\newblock \showarticletitle{Translation-based Recommendation}. In
  \bibinfo{booktitle}{{\em RecSys}}. ACM.
\newblock


\bibitem[\protect\citeauthoryear{He and McAuley}{He and McAuley}{2016a}]%
        {he2016ups}
\bibfield{author}{\bibinfo{person}{Ruining He} {and} \bibinfo{person}{Julian
  McAuley}.} \bibinfo{year}{2016}\natexlab{a}.
\newblock \showarticletitle{Ups and downs: Modeling the visual evolution of
  fashion trends with one-class collaborative filtering}. In
  \bibinfo{booktitle}{{\em WWW}}.
\newblock


\bibitem[\protect\citeauthoryear{He and McAuley}{He and McAuley}{2016b}]%
        {he2016vbpr}
\bibfield{author}{\bibinfo{person}{Ruining He} {and} \bibinfo{person}{Julian
  McAuley}.} \bibinfo{year}{2016}\natexlab{b}.
\newblock \showarticletitle{VBPR: Visual Bayesian Personalized Ranking from
  Implicit Feedback}. In \bibinfo{booktitle}{{\em AAAI}}.
\newblock


\bibitem[\protect\citeauthoryear{Lin, Liu, Sun, Liu, and Zhu}{Lin
  et~al\mbox{.}}{2015}]%
        {lin2015learning}
\bibfield{author}{\bibinfo{person}{Yankai Lin}, \bibinfo{person}{Zhiyuan Liu},
  \bibinfo{person}{Maosong Sun}, \bibinfo{person}{Yang Liu}, {and}
  \bibinfo{person}{Xuan Zhu}.} \bibinfo{year}{2015}\natexlab{}.
\newblock \showarticletitle{Learning Entity and Relation Embeddings for
  Knowledge Graph Completion.}. In \bibinfo{booktitle}{{\em AAAI}}.
\newblock


\bibitem[\protect\citeauthoryear{McAuley and Leskovec}{McAuley and
  Leskovec}{2013}]%
        {mcauley2013hidden}
\bibfield{author}{\bibinfo{person}{Julian McAuley} {and} \bibinfo{person}{Jure
  Leskovec}.} \bibinfo{year}{2013}\natexlab{}.
\newblock \showarticletitle{Hidden factors and hidden topics: understanding
  rating dimensions with review text}. In \bibinfo{booktitle}{{\em RecSys}}.
  \bibinfo{pages}{165--172}.
\newblock


\bibitem[\protect\citeauthoryear{Rendle, Freudenthaler, Gantner, and
  Schmidt-Thieme}{Rendle et~al\mbox{.}}{2009}]%
        {bpr}
\bibfield{author}{\bibinfo{person}{Steffen Rendle}, \bibinfo{person}{C.
  Freudenthaler}, \bibinfo{person}{Zeno Gantner}, {and} \bibinfo{person}{Lars
  Schmidt-Thieme}.} \bibinfo{year}{2009}\natexlab{}.
\newblock \showarticletitle{BPR: Bayesian personalized ranking from implicit
  feedback}. In \bibinfo{booktitle}{{\em UAI}}.
\newblock


\bibitem[\protect\citeauthoryear{Trewin}{Trewin}{2000}]%
        {trewin2000knowledge}
\bibfield{author}{\bibinfo{person}{Shari Trewin}.}
  \bibinfo{year}{2000}\natexlab{}.
\newblock \showarticletitle{Knowledge-based recommender systems}.
\newblock \bibinfo{journal}{{\em Encyclopedia of library and information
  science\/}} \bibinfo{volume}{69}, \bibinfo{number}{Supplement 32}
  (\bibinfo{year}{2000}), \bibinfo{pages}{180}.
\newblock


\bibitem[\protect\citeauthoryear{Wang, Zhang, Feng, and Chen}{Wang
  et~al\mbox{.}}{2014}]%
        {wang2014knowledge}
\bibfield{author}{\bibinfo{person}{Zhen Wang}, \bibinfo{person}{Jianwen Zhang},
  \bibinfo{person}{Jianlin Feng}, {and} \bibinfo{person}{Zheng Chen}.}
  \bibinfo{year}{2014}\natexlab{}.
\newblock \showarticletitle{Knowledge Graph Embedding by Translating on
  Hyperplanes}. In \bibinfo{booktitle}{{\em AAAI}}, Vol.~\bibinfo{volume}{14}.
  \bibinfo{pages}{1112--1119}.
\newblock


\bibitem[\protect\citeauthoryear{Zhang, Yuan, Lian, Xie, and Ma}{Zhang
  et~al\mbox{.}}{2016}]%
        {zhang2016collaborativekdd}
\bibfield{author}{\bibinfo{person}{Fuzheng Zhang},
  \bibinfo{person}{Nicholas~Jing Yuan}, \bibinfo{person}{Defu Lian},
  \bibinfo{person}{Xing Xie}, {and} \bibinfo{person}{Wei-Ying Ma}.}
  \bibinfo{year}{2016}\natexlab{}.
\newblock \showarticletitle{Collaborative Knowledge Base Embedding for
  Recommender Systems}. In \bibinfo{booktitle}{{\em KDD}}.
\newblock


\bibitem[\protect\citeauthoryear{Zhang, Ai, Chen, and Croft}{Zhang
  et~al\mbox{.}}{2017}]%
        {zhang2017joint}
\bibfield{author}{\bibinfo{person}{Y. Zhang}, \bibinfo{person}{Q. Ai},
  \bibinfo{person}{X. Chen}, {and} \bibinfo{person}{W.~B. Croft}.}
  \bibinfo{year}{2017}\natexlab{}.
\newblock \showarticletitle{Joint representation learning for top-n
  recommendation with heterogeneous information sources}. In
  \bibinfo{booktitle}{{\em CIKM}}.
\newblock


\bibitem[\protect\citeauthoryear{Zheng, Noroozi, and Yu}{Zheng
  et~al\mbox{.}}{2017}]%
        {zheng2017joint}
\bibfield{author}{\bibinfo{person}{Lei Zheng}, \bibinfo{person}{Vahid Noroozi},
  {and} \bibinfo{person}{Philip~S Yu}.} \bibinfo{year}{2017}\natexlab{}.
\newblock \showarticletitle{Joint deep modeling of users and items using
  reviews for recommendation}. In \bibinfo{booktitle}{{\em WSDM}}.
\newblock


\end{thebibliography}

\end{document}